\documentstyle[11pt]{article}

\input{psfig.sty}

\title{Neighborhood preferences in random matching problems}

\author{G. Parisi\footnote{giorgio.parisi@roma1.infn.it}\ \ and M. Rati\'eville\footnote{matthieu.ratieville@roma1.infn.it} \\{\normalsize Dipartimento di Fisica, INFM and INFN, Universit\`a di Roma 1 {\em La Sapienza}} \\{\normalsize P.le A. Moro, 2 -- 00185 Roma (Italy)}}

\begin{document}

\maketitle

\begin{abstract}

We consider a class of random matching problems where the distance between two points has a probability law which, for a small distance $l$, goes like $l^r$. In the framework of the cavity method, in the limit of an infinite number of points, we derive equations for $p_k$, the probability for some given point to be matched to its $k$-th nearest neighbor in the optimal configuration.  These equations are solved in two limiting cases : $r=0$ --- where we recover $p_k=1/2^k$, as numerically conjectured by Houdayer {\sl et al.} and recently rigorously proved by Aldous --- and $r \rightarrow +\infty$. For $0<r<+\infty$, we are not able to solve the equations analytically, but we compute the leading behavior of $p_k$ for large $k$.  

\end{abstract}

\section{Introduction}

Let us consider a set of $N$ points $i=1,\ldots N$, $N$ even, and 'distances' between them $l_{ij}=l_{ji}$. A matching is a partition of this set into $N/2$ pairs. The 'length' of such a matching is

\begin{equation}
L=\sum_{pair \in matching} l_{pair}
\end{equation}

We focus on the properties of the  minimal length matching. If the $l_{ij}$ are random variables, it is known that in many cases, the length of the optimal configuration and other quantities converge with probability 1 to their average value when $N \rightarrow +\infty$. A lot of different probability distributions of the $l_{ij}$ can be considered. In this note, we deal with the case where the $l_{ij}$ are independent and identically distributed random variables, with a law $ \rho(l) $ defined on $ [0,+\infty[ $. As noted in \cite{vanmez}, in the infinite $N$ limit, the only relevant feature of $\rho$ is its behavior around 0, say

\begin{equation}
\label{rho}
\rho (l) \mathop{\sim}_{l \rightarrow 0} \frac{l^r}{r!}
\end{equation}

In the following we stick to the tradition that $r$ is an integer, which simplifies results though it should not be difficult to generalize our computations to a non integer $r$. The thermodynamic limit of this model has been studied for a long time from the point of view of the statistical physics of disordered systems. The replica method yielded predictions for the mean length and the distribution of the lengths of occupied links in the optimal configuration \cite{mezpar2}. This was shown to be equivalent to a cavity approach \cite{mezpar}.

Numerical works checked the validity of the results obtained with these techniques \cite{brunetti,houdayer}, and dealt with another quantity, the probability $p_k$ for some given point to be connected to its $k$-th nearest neighbor in the optimal matching. Houdayer {\sl et al.} \cite{houdayer} conjectured that, in the case $r=0$, one simply has

\begin{equation}
\label{conjecture}
p_k=\frac{1}{2^k}
\end{equation}

Recently, Aldous \cite{aldous} confirmed the mentioned predictions by rigorous proof in the case $r=0$, and gave heuristic indications to generalize his method to an arbitrary $r$.

In this paper, we show that insight on $p_k$ can also be gained through a cavity approach, and in particular we recover (\ref{conjecture}). It gives further evidence that this method, as well as the replica one, exactly describes the matching problem, though on non rigorous grounds.
 
The paper is organized as follows. In section \ref{cavity}, given any $r$ we derive cavity equations for $p_k$ and comment on their similarity with Aldous's ones \cite{aldous}. In section \ref{r=0} we check that in the case $r=0$ we recover (\ref{conjecture}). In section \ref{rquelconque} we derive, for $0<r<+\infty$, the leading behavior of $p_k$ for large $k$. In section \ref{r=infty} we compute $p_k$ in the limit $r \rightarrow +\infty$. Section \ref{conclusion} is the conclusion. All important results are supported by numerical simulations.

\section{The cavity prediction for $p_k$}
\label{cavity}

The cavity equations at finite temperature have been derived in \cite{mezpar} (see also \cite{houches}). Let us briefly recall the essential steps --- basically we reproduce \cite{mez}. One writes down a partition function for the matching problem

\begin{equation}
Z=\sum_{matchings} \exp \left( -\beta N^\frac{1}{r+1} L_{matching} \right)
\end{equation}
where the scaling factor $N^\frac{1}{r+1}$ insures that the free energy $F=-(1/\beta) \ln Z$ is an extensive quantity.

A tractable representation of this partition function can be borrowed from the theory of polymers \cite{pgg}. One introduces on each site a $m$ dimensional vector spin ${\bf S}_i=(S_i^1,\ldots S_i^m)$ normalized by ${\bf S}_i^2=m$. Let $d\mu$ be the integration measure on the corresponding sphere. If we define $T_{ij}=\exp \left( -\beta N^{1/(r+1)} l_{ij} \right)$, one can check that

\begin{equation}
Z=\lim_{m \rightarrow 0} \int \prod_i S_i^1 d\mu({\bf S}_i) \exp \left( \sum_{i<j} T_{ij} {\bf S}_i.{\bf S}_j \right)
\end{equation}
thanks to the following property

\begin{equation}
\label{loi}
\mathop{\lim}_{m \rightarrow 0} \int d\mu({\bf S}_i) S_i^{\alpha_1} S_i^{\alpha_2} \cdots S_i^{\alpha_p}=0
\end{equation}
unless $p=2$ and $\alpha_1=\alpha_2$; in this case we have $\mathop{\lim}_{m \rightarrow 0} \int d\mu({\bf S}_i) (S_i^{\alpha})^2 =1$. The spontaneous magnetization of spin $i$ has components $m_i^\alpha=\delta_{\alpha,\,1} m_i$.

The cavity method consists in adding a new point $ i=0 $ to a system of $N$ points $ i=1,\ldots N $. The partition function of the new system is computed assuming that the correlations of the $N$ old spins can be neglected (technically speaking we suppose there is a single pure state): each spin lives in an effective field $h_i$. So one writes

\begin{equation}
Z_{N+1}=\int \prod_{i=0}^N S_i^1 d\mu({\bf S}_i) \exp \left(\sum_{i=1}^N h_i S_i^1 + \sum_{i=1}^N T_{0i} {\bf S}_0.{\bf S}_i + h_0 S_0^1 \right)
\end{equation}

This is easily computed, thanks to (\ref{loi}), and one gets by differentiation of $\ln Z$ with respect to $h_0$

\begin{equation}
\label{cavite}
m_0=\frac{1}{\sum_{i=1}^{N} T_{0i}m_{i}}
\end{equation}
where we used the fact that $ m_i $, the magnetization of site $ i $ {\sl before} the addition of the new point, is nothing but $1/h_i$.

Note that in principle $N$ should be kept even, but it has no incidence on our results in the infinite $N$ limit. The occupation number of link $ 0-i $ reads

\begin{equation}
\label{occup}
n_{0i}=m_0 T_{0i} m_i
\end{equation}

We are interested in the zero temperature limit $ \beta\rightarrow\infty $, for in this limit only the optimal configuration contributes. The probability distribution of the magnetizations in this limit has already been derived for another optimization problem, the travelling salesman problem \cite{krmez}. We follow the same lines. It is useful to set

\begin{equation}
\label{defphi}
m_i=e^{\beta\varphi_i}
\end{equation}

Let us then define

\begin{equation}
\chi_i=N^\frac{1}{r+1} l_{0i}-\varphi_i \; \; \; \; \; \; \; i=1, \ldots N
\end{equation}

At zero temperature, equation (\ref{cavite}) reduces to

\begin{equation}
\label{min}
\varphi_0=\min_{i=1,\ldots N} \chi_i
\end{equation}
and (\ref{occup}) yields 

\begin{equation}
\label{occupation}
n_{0i}=\delta_{i,\,i_{min}}
\end{equation}
where $ i_{min} $ is the indice attaining the minimum in (\ref{min}).
 
In the thermodynamic limit, the $ \varphi_i $ are uncorrelated and all have the same probability law $ P(\varphi) $ over the distribution of links. The $ \chi_i $ are also independent random variables. Their common probability law is 

\begin{eqnarray}
\Pi(\chi) & = & \int_0^{+\infty} dl \int_{-\infty}^{+\infty} d\varphi \rho(l) P(\varphi) \, \delta(N^\frac{1}{r+1} l-\varphi-\chi)
\\\nonumber & = & \frac{1}{N^\frac{1}{r+1}}\int_{0}^{+\infty}dl\, \rho \left( \frac{l}{N^\frac{1}{r+1}} \right) P(l-\chi)
\\\nonumber & \sim & \frac{1}{N}\int_{0}^{+\infty} dl\, \frac{l^r}{r!} P(l-\chi) \;\;\;\;\;\;\;\;\;\;\;\ \mbox{thanks to (\ref{rho})}
\end{eqnarray}

Equation (\ref{min}) yields a self-consistency equation for $P(\varphi)$

\begin{equation}
\label{eqp}
P(\varphi)=N\Pi(\varphi)\left( \int_{\varphi}^{+\infty}du\,\Pi(u) \right) ^{N-1}
\end{equation}

Let us define an auxiliary function

\begin{equation}
\label{defg}
G(\varphi)=\int_{0}^{+\infty}dl\, \frac{l^{r+1}}{(r+1)!} P(l-\varphi)
\end{equation}

We have $N\Pi(\varphi)=G'(\varphi)$, and

\begin{eqnarray}
\int_{\varphi}^{+\infty}du\,\Pi(u) & = & 1-\int_{-\infty}^{\varphi} du \, \Pi(u)
\\\nonumber & = & 1-\frac{1}{N} G(\varphi)
\end{eqnarray}
so that (\ref{eqp}) reads

\begin{equation}
P(\varphi)=G'(\varphi)e^{-G(\varphi)}
\end{equation}

Plugging this into (\ref{defg}) we get an integral equation for $G$

\begin{equation}
\label{eqG}
G(\varphi)=\int_{-\varphi}^{+\infty} \frac{(t+\varphi)^r}{r!} e^{-G(t)} dt
\end{equation}

We now proceed to the computation of the probability $p_k$. Point 0 to be connected to its {\em k}-th nearest neighbor in the minimum matching means that two conditions simultaneously hold:
\begin{itemize}
\item[(i)] 0 is connected to some given point $i_0$ ($N$ possible choices, e.g. $i_0=1$)
\item[(ii)] there are $ k-1 $ points ($C_{N-1}^{k-1}\sim N^{k-1}/(k-1)!$ possible choices, e.g. $i=2,\ldots k$ ) whose distances to 0 is smaller than $l_{0i_0}$.
\end{itemize}

So the probability of this event reads

\begin{eqnarray}
\label{pk}
p_k = \frac{N N^{k-1}}{(k-1)!} \left[ \prod_{i=1}^{N} \int_{-\infty}^{+\infty} \! d\varphi_i \, P(\varphi_i) \int_0^{+\infty} d\l_{0i} \, \rho (l_{0i}) \right] \left[ \prod_{i=2}^{k} \Theta(l_{01}-l_{0i}) \right] \\ \nonumber \times \left[ \prod_{i=k+1}^{N} \Theta(l_{0i}-l_{01}) \right] \left[ \prod_{i=2}^{N} \Theta \left( \varphi_1-\varphi_i+N^\frac{1}{r+1} (l_{0i}-l_{01}) \right) \right]
\end{eqnarray}
where the Heaviside functions $\Theta$ inforce conditions (i) and (ii).

Let us define $\tilde{P}$ as the following primitive of $P$

\begin{equation}
\label{defptilde}
{\tilde P}(\varphi)=\int_{-\infty}^{\varphi} du\, P(u) \, =1-e^{-G(\varphi)}
\end{equation}

We can then perform the integration with respect to the $\varphi_i, \, i=2,\ldots,N$ in (\ref{pk}). The integrals with respects to the $l_i,\, i=2, \ldots N$ factorize,

\begin{eqnarray}
p_k = \frac{N^k}{(k-1)!} \int_{-\infty}^{+\infty} d\varphi_1 \, P(\varphi_1) \, \int_0^{+\infty} dl_{01} \, \rho (l_{01}) \left[ \int_{0}^{l_{01}} dl \, \rho (l) \tilde{P} (\varphi_1+N^\frac{1}{r+1} (l-l_{01})) \right] ^{k-1} \\ \nonumber \times \left[ \int_{l_{01}}^{+\infty} dl \, \rho(l) \tilde{P} (\varphi_1+N^\frac{1}{r+1} (l-l_{01}) \right] ^{N-k}
\end{eqnarray}

After some changes of variables, it reads

\begin{eqnarray}
\label{defpk}
p_k=\frac{N^k}{(k-1)!} \int_{-\infty}^{+\infty} d\varphi \, P(\varphi) \int_0^{+\infty} \, \frac{dl}{N^\frac{1}{r+1}} \, \rho \left( \frac{l}{N^\frac{1}{r+1}} \right) \left[ \int_0^l \frac{du}{N^\frac{1}{r+1}} \, \rho \left( \frac{l-u}{N^\frac{1}{r+1}} \right) \tilde{P} (\varphi-u) \right] ^{k-1} \\ \nonumber \times \left[ \frac{1}{N^\frac{1}{r+1}} \int_0^{+\infty} \frac{du}{N^\frac{1}{r+1}} \, \rho \left( \frac{l+u}{N^\frac{1}{r+1}} \right) \tilde{P} (\varphi+u) \right] ^{N-k}
\end{eqnarray}

Simplification of the first line above is straightforward using (\ref{rho}). The second line needs some little work. It can be written

\begin{equation}
\label{truc}
\left[ \int_0^{+\infty} dv \, \rho \left( v+\frac{l}{N^\frac{1}{r+1}} \right) + \frac{1}{N^\frac{1}{r+1}} \int_{0}^{+\infty} du \, \rho \left( \frac{l+u}{N^\frac{1}{r+1}} \right) [\tilde{P}(\varphi+u)-1] \right] ^{N-k}
\end{equation}

As $\rho \left( v+\frac{l}{N^\frac{1}{r+1}} \right)= \rho (v)+ \frac{l}{N^\frac{1}{r+1}} \rho'(v) + \cdots + \frac{l^{r+1}}{(r+1)! N} \rho^{(r+1)}(v) + \cdots$, the first term in the bracket reads

\begin{eqnarray}
1 \! \! \! \! \! &+& \! \! \! \! \! \frac{l}{N^\frac{1}{r+1}} [\rho (+ \infty)-\rho(0)] +\cdots + \frac{l^{r+1}}{(r+1)! N} [ \rho^{(r)}(+\infty)-\rho^{(r)}(0)] + \cdots
\\\nonumber  &=&  1+0+\cdots+\frac{l^{r+1}}{(r+1)! N} [0-1]+ \cdots
\\ \nonumber  &=&  1-\frac{l^{r+1}}{(r+1)! N}+\cdots
\end{eqnarray}
whereas the second term is simplified once again using (\ref{rho}). We get that expression (\ref{truc}) tends to

\begin{equation}
\exp \left( -\frac{l^{r+1}}{(r+1)!} + \int_{0}^{+\infty} du \, \frac{(l+u)^r}{r!} [\tilde{P} (\varphi+u)-1] \right)
\end{equation}
and finally we find for $p_k$

\begin{eqnarray}
\label{p_k}
p_k=\frac{1}{(k-1)!} \int_{-\infty}^{+\infty} d\varphi \, P(\varphi) \int_0^{+\infty} dl \, \frac{l^r}{r!} \left[ \int_0^l du \, \frac{(l-u)^r}{r!} \tilde{P} (\varphi-u) \right]^{k-1} \\ \nonumber \times \exp \left( -\frac{l^{r+1}}{(r+1)!} +\int_0^{+\infty} du \, \frac{(l+u)^r}{r!} [\tilde{P}(\varphi+u)-1] \right) 
\end{eqnarray}

For some computations we may find easier to use a different formulation of the same result. Let us define the generating function of the $p_k$

\begin{equation}
\label{defH}
H(\varepsilon) = \sum_{k=1}^{+\infty} p_k \varepsilon^{k-1}
\end{equation}

Then $H(\varepsilon)$ is given by

\begin{eqnarray}
\label{H}
H(\varepsilon)=\int_{-\infty}^{+\infty} d\varphi P(\varphi) \int_0^{+\infty} dl\, \frac{l^r}{r!} \exp \left( -\frac{l^{r+1}}{(r+1)!} + \int_0^{+\infty} du \, \frac{(l+u)^r}{r!} [\tilde{P}(\varphi+u)-1] \right. \\ \nonumber + \left. \varepsilon  \int_0^l du \frac{(l-u)^r}{r!} \tilde{P} (\varphi-u) \right)
\end{eqnarray}

Let us comment on our solution. It involves as a basic ingredient the function $G$ determined by equation (\ref{eqG}), the other quantities being expressed in terms of $G$. This equation is actually well known, for it also appears in the solution with replicas (see \cite{mezpar2}, eq. (22)), up to normalization conventions. The advantage is that, within the cavity approach, $G$ is given a simple probabilistic interpretation in terms of the occupation numbers. This enables us to easily compute elaborate quantities such as $p_k$, which is hardly tractable with replicas.

Aldous notices that his method also yields equation (\ref{eqG}) (see section 5.1 of \cite{aldous}), but we would like to stress that his approach and ours actually meet at an earlier stage: indeed our essential equation is not (\ref{eqG}) but rather the couple (\ref{min})-(\ref{occupation}); the rest, including the introduction of $G$, is mere calculus. Roughly speaking, Aldous's method consists in linking the $N \rightarrow +\infty$ limit of the matching problem to a matching on an infinite tree. Let us concentrate on case $r=0$. Such matching is constructed introducing i.i.d. r.v. $X$ --- related to the root --- and $(X_i)_{i \geq 1}$ --- related to its children ---, and the points $(\xi_i)_{i \geq 1}$ of a Poisson process of rate 1 (see section 3.3 of \cite{aldous}). These quantities satisfy

\begin{equation}
\label{ald}
X=\min_{i \geq 1} (\xi_i - X_i)
\end{equation}
and the root of the infinite tree is matched to its child attaining the minimum.

This is strongly reminiscent of (\ref{min})-(\ref{occupation}). More precisely, if one relabels the $l_{0i}$ so that they be sorted in increasing order, and consistently relabel the $\varphi_i$, the probability law of $(N l_{0i})_{i \geq 1}$ tends to a Poisson process of rate 1. Thus (\ref{min}) becomes identical to (\ref{ald}), provided we identify $X_i$ and $\varphi_i$.

Indeed, if we also take into account Aldous's heuristic considerations, similarities can be found even before (\ref{min}): in section 5.3 of \cite{aldous}, Aldous gives indications for a possible extension of his method to the finite temperature problem. Using the same kind of r.v. as in (\ref{ald}) he writes down the equation

\begin{equation}
\label{ald2}
X=\left( \sum_{i=1}^{+\infty} e^{-\lambda \xi_i} X_i \right) ^{-1}
\end{equation}
which can be identified with the cavity equation (\ref{cavite}). Of course, the same way we derive (\ref{min}) from (\ref{cavite}) using trick (\ref{defphi}), Aldous observes that (\ref{ald2}) is consistent with (\ref{ald}) in the $\lambda \rightarrow +\infty$ limit.

Such formal similarities are striking but we are unable to account for them. Whether some further study can lead to more a rigorous basis for the cavity method is not clear, because we define the $m_i$,$\, \varphi_i$ through the somewhat exotic procedure of taking the $m \rightarrow 0$ limit of a $m$-spin model.

\section{The case $r=0$}
\label{r=0}

For $r=0$, equation (\ref{eqG}) is easily solved. One finds

\begin{equation}
G(\varphi)=\ln(1+e^{\varphi})
\end{equation}

It turns out that the computation of $H(\varepsilon)$ is simpler than the direct computation of $p_k$. Here (\ref{H}) reduces to

\begin{eqnarray}
H(\varepsilon) & = & \int_{-\infty}^{+\infty} \! d\varphi P(\varphi) \exp \left( \int_0^{+\infty} \! du [\tilde{P}(\varphi+u)-1] \right) \! \int_0^{+\infty} \! dl e^{-l}  \exp \left(\varepsilon \! \int_0^l \! du \tilde{P} (\varphi-u) \right)
\\ \nonumber
& = & \int_{-\infty}^{+\infty} d\varphi \frac{e^\varphi}{(1+e^\varphi)^2} \frac{e^\varphi}{1+e^\varphi}  \int_0^{+\infty} dl \, e^{-l} \exp \left( \varepsilon \ln \frac{1+e^\varphi}{1+e^{\varphi-l}} \right)
\\ \nonumber
& = & \int_0^{+\infty} dy \frac{y}{(1+y)^3} \int_0^1 dx \, \left( \frac{1+y}{1+yx} \right) ^\varepsilon 
\\ \nonumber
& = & \int_0^{+\infty} dy \frac{y}{(1+y)^3} \frac{1+y-(1+y)^\varepsilon}{(1-\varepsilon) y}
\end{eqnarray}

Provided $\varepsilon <2$ we finally get

\begin{equation}
H(\varepsilon)=\frac{1}{2-\varepsilon}
\end{equation}
which immediately yields the expected result (\ref{conjecture}).

\section{The case $0 < r < +\infty$ }
\label{rquelconque}

In this case, we are not able to find an explicit solution of (\ref{eqG}). To check the validity of our approach, we solved this equation numerically by iterations for $r=1,2,3$ and computed $p_k$ using (\ref{p_k}). The results compare very well with the estimates of $p_k$ we got numerically by averaging over 40000 random instances of size $N=200$ (see table \ref{table}).

\begin{table}
\begin{tabular}{|l|c|c|c|c|c|c|c|c|}
\cline{2-7} \cline{9-9}
\multicolumn{1}{c}{} & \multicolumn{2}{|c|}{$r=1$} & \multicolumn{2}{c|}{$r=2$} & \multicolumn{2}{c|}{$r=3$} & & \multicolumn{1}{|c|}{$r\rightarrow +\infty$} \\\cline{2-7} \cline{9-9}
 \multicolumn{1}{c|}{} & (1) &
 (2) & (1) & (2) & (1) & (2) & & (2)
\\\cline{1-7} \cline{9-9} $p_1$ & 0.52891(15) & 0.52788 & 0.53853(15) & 0.53865 & 0.54308(15) & 0.54320 & & 0.55523
\\\cline{1-7} \cline{9-9} $p_2$ & 0.22790(15) & 0.22816 & 0.21706(15) & 0.21789 & 0.21118(15) & 0.21192 && 0.19329
\\\cline{1-7} \cline{9-9} $p_3$ & 0.11215(11) & 0.11239 & 0.10634(11) & 0.10702 & 0.10318(11) & 0.10393 && 0.09399
\\\cline{1-7} \cline{9-9} $p_4$ & 0.05875(8) & 0.05882 & 0.05719(8) & 0.05742 & 0.05628(8) &0.05648 && 0.05278
\\\cline{1-7} \cline{9-9} $p_5$ & 0.03185(6) & 0.03188 & 0.03234(6) & 0.03243 & 0.03247(6) & 0.03261 && 0.03222
\\\cline{1-7} \cline{9-9} $p_6$ & 0.01753(5) & 0.01766 & 0.01881(5) & 0.01895 & 0.01946(5) & 0.01960 && 0.02079
\\\cline{1-7} \cline{9-9} $p_7$ & 0.00986(4) & 0.00992 & 0.01126(4) & 0.01135 & 0.01208(4) & 0.01212 && 0.01396
\\\cline{1-7} \cline{9-9} $p_8$ & 0.00556(3) & 0.00564 & 0.00689(3) & 0.00692 & 0.00763(3) & 0.00767 && 0.00967
\\\cline{1-7} \cline{9-9} $p_9$ & 0.00317(2) & 0.00323 & 0.00424(2) & 0.00429 & 0.00491(3) & 0.00494 && 0.00687
\\\cline{1-7} \cline{9-9} $p_{10}$ & 0.00179(2) & 0.00186 & 0.00264(2) & 0.00269 & 0.00320(2) & 0.00323 && 0.00497
\\\cline{1-7} \cline{9-9}
\end{tabular}
\caption{$p_k$ computed (1) by averaging over 40000 samples of size N=200, (2) by numerical integration of our equations}
\label{table}
\end{table}

Still we wanted to see if one could state as a simple law as (\ref{conjecture}), at least asymptotically for large $k$. Rewrite (\ref{p_k}) as

\begin{equation}
\label{pktoujours}
p_{k+1}=\frac{1}{k!} \int_{-\infty}^{+\infty} d\varphi \int_0^{+\infty}  dl \, G'(\varphi) \frac{l^r}{r!} e^{A(k,\varphi,l)}
\end{equation}
where

\begin{eqnarray}
\label{premierA}
A(k,\varphi,l)= -G(\varphi) -\frac{l^{r+1}}{(r+1)!} + \int_0^{+\infty} du \frac{(l + u)^r}{r!} \left[ \tilde{P} (\varphi + u)-1 \right]
\\ \nonumber +k \ln \left[ \int_0^{l} du \frac{(l-u)^r}{r!} \tilde{P} (\varphi -u) \right]
\end{eqnarray}

To get the asymptotic behavior of such an integral, we focus on the $k$-dependent maximum of $A(k,\varphi,l)$ with respect to $(\varphi,l)$. In the following we exhibit a stationary point and, assuming it is the absolute maximum, carry out the computation of the leading behavior of $p_k$.

We start deriving an asymptotic expansion of $A(k,\varphi,l)$ for large $k$, $\varphi$ and $l$, under the constraints

\begin{eqnarray}
\label{hypo}
1<C_1<&l/\varphi&<C_2<+\infty
\\\label{hypo2}
\varphi&>&k^\nu
\end{eqnarray}
for some arbitrary but fixed $C_1$,$C_2$ and $\nu>0$. We now deal in turn with each term in (\ref{premierA}).

\begin{itemize}

\item

If we write (\ref{defg}) as follows

\begin{equation}
G(\varphi)=\int_{-\varphi}^{+\infty}dt\, \frac{(t+\varphi)^{r+1}}{(r+1)!} P(t)
\end{equation}
then expand $(t+\varphi)^{r+1}$, and split the integral, we get

\begin{equation}
G(\varphi) = \sum_{m=0}^{r+1} \frac{\varphi^m}{m!} \int_{-\infty}^{+\infty} dt \frac{t^{r+1-m}}{(r+1-m)!} P(t) - \sum_{m=0}^{r+1} \frac{\varphi^m}{m!} \int_{-\infty}^{-\varphi} dt \frac{t^{r+1-m}}{(r+1-m)!} P(t)
\end{equation}

The terms of the second sum above all go to 0 (it is easy to derive from (\ref{eqG}) that $P$ goes to zero faster than any power both in $+\infty$ and $-\infty$). So

\begin{equation}
\label{asymptG}
G(\varphi)=\frac{\varphi^{r+1}}{(r+1)!} + \sum_{m=0}^{r} a_m \varphi^m + o(1)
\end{equation}
where the coefficients $a_m$ are essentially moments of the distribution $P$

\begin{equation}
a_m=\frac{1}{m!} \int_{-\infty}^{+\infty} dt \frac{t^{r+1-m}}{(r+1-m)!} P(t)
\end{equation}

\item Using the definition (\ref{defptilde}) of $\tilde{P}$, the third term in (\ref{premierA}) can be written

\begin{equation}
-\sum_{m=0}^r \frac{(l-\varphi)^m}{m!} \int_{\varphi}^{+\infty} dt \frac{t^{r-m}}{(r-m)!} e^{-G(t)}
\end{equation}

All terms go to zero --- it is essential here to be able to compare $\varphi$ and $l$; that is why we set (\ref{hypo}).

\item Finally, the argument of the $\ln$ in the fourth term of (\ref{premierA}) can be integrated by parts, giving

\begin{eqnarray}
&& k \ln \left[ \frac{l^{r+1}}{(r+1)!} \left[ 1-e^{-G(\varphi)} \right] - \int_{\varphi-l}^{\varphi} dt \frac{(t+l-\varphi)^{r+1}}{(r+1)!} P(t) \right]
\\\label{intermediaire} & =& k \ln \left[ \frac{l^{r+1}}{(r+1)!} + o(1) - \frac{(l-\varphi)^{r+1}}{(r+1)!} - \sum_{m=0}^{r} (l-\varphi)^m a_m + o(1) \right]
\\\nonumber &=& k \ln \left[ \frac{l^{r+1}}{(r+1)!} - \frac{(l-\varphi)^{r+1}}{r+1)!} - \sum_{m=0}^{r} (l-\varphi)^m a_m \right] +o(1)
\end{eqnarray}
because the $o(1)$ terms at line (\ref{intermediaire}) go to 0 faster than any power of $k$ thanks to (\ref{hypo})-(\ref{hypo2}). 

\end{itemize}

Collecting all the pieces, we get

\begin{eqnarray}
\label{A}
\lefteqn{A(k,\varphi,l)=-\frac{l^{r+1}}{(r+1)!} -\frac{\varphi^{r+1}}{(r+1)!} - \sum_{m=0}^{r} a_m \varphi^m} \\ \nonumber & & + k \ln \left[ \frac{l^{r+1}-(l-\varphi)^{r+1}}{(r+1)!}- \sum_{m=0}^r a_m (l-\varphi)^m \right]+o(1)
\end{eqnarray}

Dropping the $o(1)$ term, the stationarity equations read

\begin{eqnarray}
-\frac{\varphi^r}{r!}- \sum_{m=0}^{r-1} (m+1) a_{m+1} \varphi^m  + k \frac{\frac{(l-\varphi)^r}{r!}+\sum_{m=0}^{r-1} (m+1) a_{m+1} (l-\varphi)^m} {\frac{l^{r+1}-(l-\varphi)^{r+1}}{(r+1)!}- \sum_{m=0}^r a_m (l-\varphi)^m}=0
\\ \nonumber
-\frac{l^r}{r!} + k \frac{\frac{l^r-(l-\varphi)^r}{r!} - \sum_{m=0}^{r-1} (m+1) a_{m+1} (l-\varphi)^m} {\frac{l^{r+1}-(l-\varphi)^{r+1}}{(r+1)!}- \sum_{m=0}^r a_m (l-\varphi)^m}=0
\end{eqnarray}

Let us call $(\varphi_k,l_k)$ the solution of this system. We can see that both $\varphi_k$ and $l_k$ scale as $k^{1/(r+1)}$, which is consistent with hypotheses (\ref{hypo}) and (\ref{hypo2}). If we write

\begin{eqnarray}
\label{lkphik}
\ \varphi_k & = & k^\frac{1}{r+1} \hat{\varphi}_k
\\ \nonumber l_k & = & k^\frac{1}{r+1} \hat{l}_k
\end{eqnarray}
the system reads

\begin{eqnarray}
\label{system}
\lefteqn{-\frac{\hat{\varphi}_k^r}{r!}- \sum_{m=0}^{r-1} k^\frac{m-r}{r+1} (m+1) a_{m+1} \hat{\varphi}_k^m} \\ \nonumber & & \; \; \; \;\; \;\; \;\; \; + \frac{\frac{(\hat{l}_k-\hat{\varphi}_k)^r}{r!}+\sum_{m=0}^{r-1} k^\frac{m-r}{r+1} (m+1) a_{m+1} (\hat{l}_k-\hat{\varphi}_k)^m} {\frac{\hat{l}_k^{r+1}-(\hat{l}_k-\hat{\varphi}_k)^{r+1}}{(r+1)!}- \sum_{m=0}^r k^{\frac{m}{r+1}-1} a_m (\hat{l}_k-\hat{\varphi}_k)^m} = 0
\\ \nonumber
\lefteqn{-\frac{\hat{l}_k^r}{r!} + \frac{\frac{\hat{l}_k^r-(\hat{l}_k-\hat{\varphi}_k)^r}{r!} - \sum_{m=0}^{r-1} k^\frac{m-r}{r+1} (m+1) a_{m+1} (\hat{l}_k-\hat{\varphi}_k)^m} {\frac{\hat{l}_k^{r+1}-(\hat{l}_k-\hat{\varphi}_k)^{r+1}}{(r+1)!}- \sum_{m=0}^r k^{\frac{m}{r+1}-1} a_m (\hat{l}_k-\hat{\varphi}_k)^m} = 0}
\end{eqnarray}

We face a perturbative problem whose small parameter is $k^{-1/(r+1)}$. Knowing the $a_m$, it is in principle possible to compute iteratively the following asymptotic expansions of $\hat{\varphi}_k$ and $\hat{l}_k$

\begin{eqnarray}
\label{expansion}
\hat{\varphi}_k & = & \hat{\varphi}_\infty+ \sum_{m=1}^{r+1} c_m k^{-\frac{m}{r+1}} + o(k^{-1})
\\ \nonumber
\hat{l}_k & = & \hat{l}_\infty+ \sum_{m=1}^{r+1} d_m k^{-\frac{m}{r+1}} + o(k^{-1})
\end{eqnarray}

Though we cannot compute all the coefficients in closed form for a generic $r$, we can give a simple procedure to determine the leading ones, $\hat{l}_\infty$ and $\hat{\varphi}_\infty$. We define $\alpha =\hat{l}_\infty/\hat{\varphi}_\infty$. Combining the two equations (\ref{system}), where one sets $k=+\infty$, yields

\begin{equation}
\label{eqpouralpha}
\frac{1}{(\alpha-1)^r}-\frac{1}{\alpha^r}=1
\end{equation}

Note that $\alpha$ must be greater than 1 to be consistent with hypothesis (\ref{hypo}). On $]1,+\infty[$ the left hand side of (\ref{eqpouralpha}) is a decreasing function of $\alpha$, from $+\infty$ down to 0, so the above equation does admit a solution in this interval. Other interesting properties of $\alpha$ are derived in appendix A. The first line of system (\ref{system}) gives

\begin{eqnarray}
\hat{\varphi}_\infty^{r+1} = (r+1)! \frac{1}{\alpha^{r+1}+1}
\end{eqnarray}
from which follow two relations we will need further down

\begin{eqnarray}
\label{usefulrelations}
\hat{\varphi}_\infty^{r+1}+\hat{l}_\infty^{r+1} & = & (r+1)!
\\ \nonumber \hat{l}_\infty^{r+1}-(\hat{l}_\infty-\hat{\varphi}_\infty)^{r+1} & = & (r+1)! (\alpha-1)^r
\end{eqnarray}

In (\ref{pktoujours}) we perform the rescaling $\varphi \leftarrow \varphi_k \varphi$ and $l \leftarrow l_k l$, so that

\begin{equation}
p_{k+1} \sim \frac{1}{k!} \varphi_k \frac{l_k^{r+1}}{r!} \int_{-\infty}^{+\infty} d\varphi \int_0^{+\infty} dl l^r G'(\varphi_k \varphi) e^{A(k,\varphi_k \varphi, l_k l)}
\end{equation}

To get the asymptotic behavior of the integral, we expand $A(k,\varphi_k \varphi, l_k l)$ around its maximum $(\varphi=1,l=1)$, up to quadratic terms, and extend the range of integration with respect to $l$ down to $-\infty$

\begin{equation}
\label{asympt}
p_{k+1} \sim \frac{1}{k!} \varphi_k \frac{l_k^{r+1}}{r!} \frac{2\pi}{\sqrt{\det \partial^2 A}} G'(\varphi_k) e^{A(k,\varphi_k, l_k)}
\end{equation}
where $ \det \partial^2 A = \frac{\partial^2 A(\varphi_k \varphi,l_k l)}{\partial \varphi^2} \frac{\partial^2 A(\varphi_k \varphi,l_k l)}{\partial l^2} -  \left( \frac{\partial^2 A(\varphi_k \varphi,l_k l)}{\partial \varphi \partial l} \right)^2 |_{\varphi=1,l=1} $. Looking at (\ref{A}) and (\ref{lkphik}) it is clear that (\ref{expansion}) induces an expansion of $A(k,\varphi_k \varphi ,l_k l)$ in powers of $k^{1/(r+1)}$ of the following form (for finite $\varphi$ and $l$ now)

\begin{eqnarray}
\lefteqn{A(k,\varphi_k \varphi ,l_k l,k)= - k \frac{(\hat{l}_\infty l)^{r+1} + (\hat{\varphi}_\infty \varphi)^{r+1}}{(r+1)!} + \sum_{m=0}^r k^\frac{m}{r+1} f_m(\varphi,l)}
\\ \nonumber & & + k \ln \left[ k \frac{(\hat{l}_\infty l)^{r+1} -(\hat{l}_\infty l - \hat{\varphi}_\infty \varphi)^{r+1}}{(r+1)!} + \sum_{m=0}^r k^\frac{m}{r+1} g_m(\varphi,l) \right] +o(1)
\end{eqnarray}

In this, $\ln[\cdots]$ can be further expanded, yielding

\begin{eqnarray}
\label{truc2}
\lefteqn{A(k,\varphi_k \varphi,l_k l) = - k \frac{(\hat{l}_\infty l)^{r+1} + (\hat{\varphi}_\infty \varphi)^{r+1}}{(r+1)!} + k \ln k }
\\ \nonumber & & + k \ln \left[ \frac{(\hat{l}_\infty l)^{r+1} -(\hat{l}_\infty l - \hat{\varphi}_\infty \varphi)^{r+1}}{(r+1)!} \right] +\sum_{m=0}^r k^\frac{m}{r+1} s_m(\varphi,l) +o(1)
\end{eqnarray}

It follows that $\det \partial^2 A \propto k^2$, (the symbol $\propto$ stands for 'goes like \ldots \  up to some constant factor'). The prefactor is a complicated but easily computed function of $\alpha$, which we omit to write.

Equation (\ref{asymptG}) tells us that $G'(\varphi_k) \sim \varphi_k^r/r!$, so that the factor $\varphi_k l_k^{r+1}  G'(\varphi_k)$ in (\ref{asympt}) is $\propto k^2$.

Beside, using (\ref{usefulrelations}) and (\ref{truc2}) we get

\begin{equation}
A(k,\varphi_k,l_k)= -k+k \ln k + k r \ln(\alpha-1)+\sum_{m=0}^r \lambda_m k^\frac{m}{r+1} +o(1)
\end{equation} 
where $\lambda_m=s_m(1,1)$.

Dropping numerical prefactors, (\ref{asympt}) thus reads

\begin{equation}
p_{k+1} \propto \frac{1}{k!} k e^{-k} k^k  \left[ (\alpha-1)^r \right] ^k \exp \left( \sum_{m=1}^r \lambda_m k^\frac{m}{r+1} \right)
\end{equation}

Sterling formula finally yields

\begin{equation}
\label{enfin}
p_{k+1} \propto \sqrt{k} \left[ (\alpha-1)^r \right] ^k \exp \left( \sum_{m=1}^r \lambda_m k^\frac{m}{r+1} \right)
\end{equation}

We have above the leading behavior of $p_{k+1}$. It is clear that an identical formula holds for $p_k$, but with different constants $\lambda_m$.

This is certainly less appealing than the simplicity of case $r=0$. The leading behavior of $p_k$ involves more and more elementary functions as $r$ increases. It makes us understand why Houdayer {\sl et al.} \cite{houdayer} were not able to find any convincing fit of their numerical estimates of $p_k$ for $r>0$.

Note that as a consequence of (\ref{enfin}), we can state the simpler but weaker following property

\begin{equation}
\label{weaker}
\frac{\ln p_k}{k} \rightarrow r \ln (\alpha-1)
\end{equation}  

This limit is negative, as expected, increases with $r$ and goes to 0 like $1/2^r$ when $r \rightarrow +\infty$ (see appendix A for details).

To check prediction (\ref{enfin}) we carried out numerically the whole computation sketched above, including constants, in the particular case $r=1$. We got
\begin{equation}
\label{asymptr=1}
p_k \sim 0.3387 \sqrt{k} (0.6180)^k e^{-0.6014 \sqrt{k}}
\end{equation}

\begin{figure}
\centerline{\psfig{file=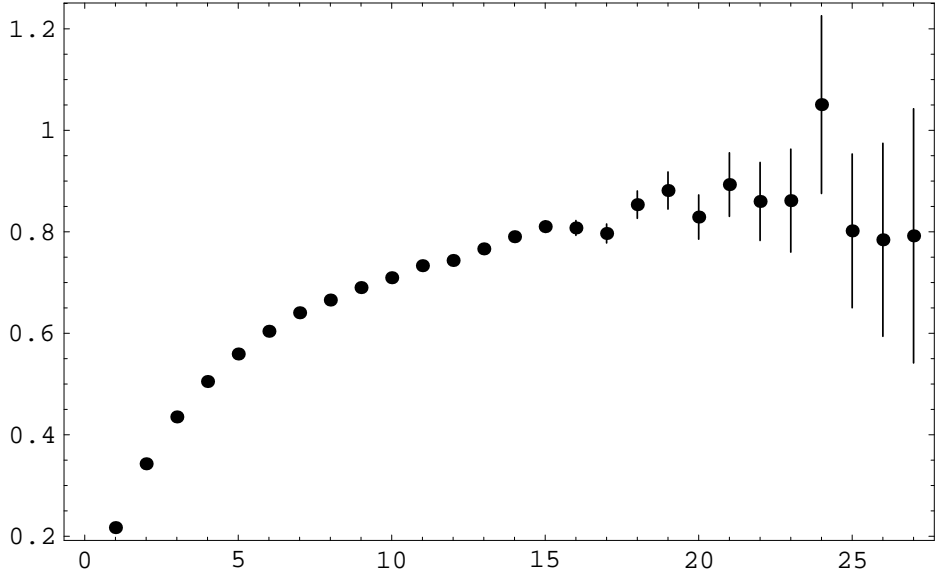}}
\caption{In the case $r=1$, plot of the ratio [asymptotic behavior of equation (\ref{asymptr=1})]/[numerical estimate of $p_k$ by averaging over 150000 samples of size $N=300$] versus $k$}
\label{figure}
\end{figure}

It is compatible with our numerical simulations (see fig. \ref{figure}).

\section{The case $r \rightarrow +\infty$}
\label{r=infty}

Our aim is to compute $p_k$ in the limit $r \rightarrow +\infty$, where new simplifications appear. We get inspiration from a similar computation performed in \cite{houdayer}. We define the following function

\begin{equation}
\hat{G}(x)=G \left[ \left( \frac{1}{2}+\frac{x}{r} \right) (rr!)^\frac{1}{r+1} \right]
\end{equation}

In terms of $\hat{G}$, (\ref{eqG}) reads

\begin{equation}
\hat{G}(x)=\int_{-r-x}^{+\infty} \left( 1+\frac{x+y}{r} \right) ^r e^{-\hat{G}(y)} dy
\end{equation}

When we send $r \rightarrow +\infty$ this equation has the following limit

\begin{equation}
\hat{G}(x)=\int_{-\infty}^{+\infty} e^{x+y} e^{-\hat{G}(y)} dy
\end{equation}
whose solution is very simple

\begin{equation}
\hat{G}(x)=e^x
\end{equation}

As in the case $r=0$, our choice is to first compute $H(\varepsilon)$. It is easy to explicitly rewrite (\ref{H}) in terms of $G$, using (\ref{eqG}) and (\ref{defptilde})

\begin{eqnarray}
\lefteqn{H(\varepsilon)= \int_{-\infty}^{+\infty} d\varphi \, G'(\varphi) e^{-G(\varphi)} \int_{-\varphi}^{+\infty} du \frac{(u+\varphi)^r}{r!} e^{-G(u)}}
\\\nonumber & & \times \exp \left[ (\varepsilon-1) \left( \frac{(u+\varphi)^{r+1}}{(r+1)!} - \int_{-u}^{\varphi} dt \frac{(t+u)^r}{r!} e^{-G(t)} \right) \right]
\end{eqnarray}

Then the changes of variables $t=(1/2+v/r)(rr!)^{1/(r+1)}$, $u=(1/2+w/r)(rr!)^{1/(r+1)}$ and $\varphi=(1/2+\psi/r)(rr!)^{1/(r+1)}$ give

\begin{eqnarray}
H(\varepsilon) &=& \int_{-\infty}^{+\infty} d\psi \, \hat{G}'(\psi) e^{-\hat{G}(\psi)} \int_{-\infty}^{+\infty} dw \, e^{w+\psi} e^{-\hat{G}(w)} 
\\\nonumber && \;\;\;\;\;\;\;\;\;\;\;\; \times \exp \left[ (\varepsilon-1) \left( e^{w+\psi} - \int_{-\infty}^\psi dv e^{v+w} e^{-\hat{G}(v)} \right) \right]
\\ \nonumber
&=& \int_{-\infty}^{+\infty} d\psi \, \frac{e^{2\psi} e^{-e^\psi}} {(1-\varepsilon) \left[ e^\psi + e^{-e^\psi} -1 \right] +1}
\\ \nonumber
&=& \int_0^{+\infty} dt \, \frac{te^{-t}}{(1-\varepsilon) \left[ t+e^{-t}-1 \right] +1}
\end{eqnarray}

Remembering (\ref{defH}), this means that

\begin{equation}
\label{pkrinf}
p_k=\int_0^{+\infty} \frac{t\,dt}{1+te^t} \left( 1-\frac{1}{t+e^{-t}} \right) ^{k-1}
\end{equation}

We are not able to do any further analytical simplification. However we can derive the asymptotic behavior of $p_k$. The integral in (\ref{pkrinf}) is dominated by the region $t \rightarrow +\infty$, so that

\begin{eqnarray}
p_k &\sim& \int_1^{+\infty} dt\, e^{-t} \left( 1-\frac{1}{t} \right) ^{k-1}
\\ \nonumber
&\sim& \sqrt{k} \int_\frac{1}{\sqrt{k}}^{+\infty} du \, e^{-\sqrt{k}u} \left( 1-\frac{1}{\sqrt{k}u} \right) ^{k-1}
\\ \nonumber
&\sim& \sqrt{k} \int_\frac{1}{\sqrt{k}}^{+\infty} du \, e^{-\frac{1}{2u^2}} e^{-\sqrt{k} \left( u+\frac{1}{u} \right)}
\end{eqnarray}

The argument of the second exponential above has its maximum in $u=1$. A saddle-point method yields

\begin{equation}
\label{eqpkrinf}
p_k \sim \sqrt{\frac{\pi}{e}}  k^{1/4} e^{-2\sqrt{k}}
\end{equation}

 The values of $p_k$ obtained by numerical integration of (\ref{pkrinf}) for $r=1,\ldots 10$ are given in table \ref{table}. Extrapolating upon comparison with results for $r=0,1,2,3$, one can see that for small $k$, say up to $k \sim 6$, $p_k$ little varies when $r$ increases. And these are the values of $k$ for which $p_k$ is significant. For larger values of $k$ the range of relative variation is more important, and diverges when $k \rightarrow +\infty$ (compare (\ref{eqpkrinf}) and (\ref{conjecture})). But then anyway $p_k$ remains very small. Beside, it seems that $p_2$, $p_3$ and $p_4$ decrease with $r$ whereas all other $p_k$ increase.  

We have not tried to see whether some of these features hold for the Euclidean random matching problem in dimension $d$ --- where the $l_{ij}$ are the usual distances between $N$ points independently and uniformly distributed in the hypercube $[0,1]^d$, which is believed to be equivalent to the uncorrelated version we study here in the limit $r=d-1 \rightarrow +\infty$ \cite{houdayer,mezpar4}. In particular, one expects $p_k$ in this limit to be the same for both versions.

\section{Conclusion}
\label{conclusion}

We have shown that the cavity method yields predictions for $p_k$ which are consistent with both rigorous results and numerical simulations.

Let us emphasize that our results also hold for the bipartite minimum matching (or assignment) problem. It has already been observed that the latter and the problem we discussed in this paper are equivalent in the $N \rightarrow +\infty$ limit, up to normalization conventions \cite{houdayer,aldous}. In appendix B we briefly show how this can be seen within our formalism.

\bigskip

\appendix

\section*{Appendix A: Derivation of some properties of $\alpha$}
\label{propalpha}

Here we write $\alpha_r$ for $\alpha$, to highlight its dependence on $r$. Let $f_r$ be the function given on $]1,+\infty[$ by

\begin{equation}
f_r(x)=\frac{1}{(x-1)^r} -\frac{1}{x^r}
\end{equation}

$f_r$ strictly decreases from $+\infty$ down to 0, and $\alpha_r$ is defined by

\begin{equation}
\label{pouralpha}
f_r(\alpha_r)=1
\end{equation}

A small computation leads to

\begin{eqnarray}
f_r(2-\frac{1}{r 2^r}) &=& 1+\frac{1}{2 r 4^r}+ o(\frac{1}{ r 4^r})
\bigskip \\\nonumber &>& 1 \; \; \; \; \;  \mbox{for large enough r}
\end{eqnarray}
and

\begin{eqnarray}
f_r(2)=1-\frac{1}{2^r}<1
\end{eqnarray}
so that, for large enough $r$, 

\begin{equation}
\label{encadre}
2-\frac{1}{r 2^r}<\alpha_r<2
\end{equation}

Beside,

\begin{equation}
f_{r+1}(\alpha_r^\frac{r}{r+1})=\frac{1}{(\alpha_r^\frac{r}{r+1} -1)^{r+1}} -\frac{1}{\alpha_r^r}
\end{equation}

One has $\alpha_r^\frac{r}{r+1}<\alpha_r$ and so $(\alpha_r^\frac{r}{r+1} -1)^{r+1}<(\alpha_r-1)^{r+1}<(\alpha_r-1)^r$ thanks to the right hand side of (\ref{encadre}). Thus

\begin{equation}
f_{r+1}(\alpha_r^\frac{r}{r+1})>f_r(\alpha_r)=1
\end{equation}

It follows that

\begin{eqnarray}
\alpha_r^\frac{r}{r+1}  &<& \alpha_{r+1}
\\\nonumber 1+\frac{1}{ \alpha_r^r} &>& 1+\frac{1}{\alpha_{r+1}^{r+1}}
\\\nonumber \mbox{i.e.} \; \;\;\;\;\;\;\;\; \frac{1}{(\alpha_r-1)^r} &>& \frac{1}{(\alpha_{r+1}-1)^{r+1}} \; \;\;\;\;\;\;\;\; \mbox{by (\ref{pouralpha})}
\end{eqnarray}

$(\alpha_r-1)^r$ increases with $r$, and goes to 1 when $r \rightarrow +\infty$, as shown by (\ref{encadre}). More precisely, it goes exponentially fast to this limit: (\ref{pouralpha}) implies

\begin{eqnarray}
\nonumber 2^r \left[ \frac{1}{(\alpha_r-1)^r}-1 \right] &=& \left( \frac{2}{\alpha_r} \right)^r \rightarrow 1 \; \;\;\;\;\;\;\;\; \mbox{by (\ref{encadre})}
\\ \mbox{i.e.} \;\;\;\;\;\;\;\;\;\;\;\; (\alpha_r-1)^r &=& 1-\frac{1}{2^r}+o(\frac{1}{2^r})
\end{eqnarray} 

\section*{Appendix B: Note on the bipartite minimum matching problem}
\label{bipartite}

The bipartite minimum matching problem can be stated as follows: take $ 2N $ points $ i=1,\ldots 2N $ divided into two subsets, $A$ ($ i=1,\ldots N $) and $B$ ($ i=N+1,\ldots 2N $) such that the distances between a point in $A$ and a point in $B$ follow the probability distribution $ \rho $ of (\ref{rho}), and allow only matchings where each pair is made of a point from A and a point from B.  One can see it as a particular case of non bipartite matching problem for the points in $A \cup B$, setting that the distances inside each subset are infinite, i.e. $ T_{ij}=0 $ for $ (i,j) \in \{ 1,\ldots N \}^2 $ or $ (i,j) \in \{ N+1,\ldots 2N \}^2 $. Then imagine that you add a new point $i=0$ to, say, subset $A$ .The cavity equation (\ref{cavite}) is slightly modified

\begin{equation}
m_0=\frac{1}{\sum_{i=N+1}^{2N} T_{0i}m_{i}}
\end{equation}
where

\begin{equation}
T_{ij}=e^{-\beta (2N)^\frac{1}{r+1} l_{0i}}
\end{equation}

Provided we relabel the $m_i$, $i\!=\!N\!+\!1, \ldots 2N$ and redefine $\beta$, we get exactly the same equation (\ref{cavite}) as in the case of the non bipartite matching problem. Indeed it is known from the replica approach \cite{mezpar3} that, for systems where the lengths $l_{ij}$ are uncorrelated random variables, the bipartite and non bipartite matching problems have the same limit when $N \rightarrow +\infty$ and differ in the $1/N$ corrections.

\end{document}